\documentclass[journal=jpclcd
,layout=twocolumn
]{achemso}

\usepackage{graphicx}
\usepackage{dcolumn}
\usepackage{bm}
\usepackage{physics}
\usepackage[version=4]{mhchem}
\usepackage{xcolor}

\def\br{{\mathbf{r}}}

\title[Sample title]{Non-equilibrium Theoretical Framework and Universal Design Principles of Oscillation-Driven Catalysis}

\author{Zhongmin Zhang}
\email{zzm@email.unc.edu}
\affiliation{Department of Chemistry,
University of North Carolina,
Chapel Hill, NC 27599-3290, U.S.A.}
\author{Zhiyue Lu}
\email{zhiyuelu@unc.edu}
\affiliation{Department of Chemistry,
University of North Carolina,
Chapel Hill, NC 27599-3290, U.S.A.}

\begin{tocentry}
\includegraphics{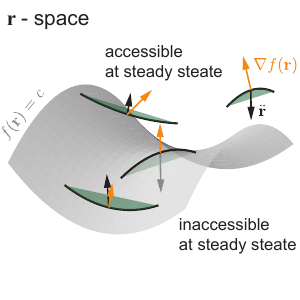}
\end{tocentry}

\begin{document}

\begin{abstract}
At stationary environmental conditions, a catalyst's reaction kinetics may be restricted by its available designs and thermodynamic laws. Thus its stationary performances may experience practical or theoretical restraints (e.g., catalysts cannot invert the spontaneous direction of a chemical reaction). However, many works have reported that if environments change rapidly, catalysts can be driven away from stationary states and exhibit anomalous performance. We present a general geometric non-equilibrium theory to explain anomalous catalytic behaviors driven by rapidly oscillating environments where stationary-environment restraints are broken. It leads to a universal design principle of novel catalysts with oscillation-pumped performances. Even though a single free energy landscape cannot describe catalytic kinetics at various environmental conditions, we propose a novel control-conjugate landscape to encode the reaction kinetics over a range of control parameters $\lambda$, inspired by the Arrhenius form. The control-conjugate landscape significantly simplifies the design principle applicable to large-amplitude environmental oscillations.
\end{abstract}

\maketitle

Traditional catalysis theory utilizes catalytic reaction pathways with reduced activation barriers to speed up both forward and backward reactions, but the catalyst itself cannot alter the spontaneous direction of the reaction. The thermodynamic driving force of a reaction can be determined either internally, namely by the free energy difference between reactants and products, or externally, namely by the external driving forces, such as the free energy dissipation from auxiliary processes accompanying the reaction. When coupled to external driving forces, a catalyst could alter the internal thermodynamic driving force and invert the spontaneous direction of a chemical reaction \cite{hill2012free,qian2021stochastic,ge2012stochastic}. Examples of externally driven catalysis can be found in light-harvesting catalysis \cite{Gratzel1983-rx} and electrolysis \cite{Bard2000-vw}). In the past decades, it has been reported that time-varying (oscillating) environments could serve as external driving forces to alter the performance of catalysts beyond their steady-state performances \cite{Ardagh2019-uj,Gathmann2022-rn,Ardagh2020-iu,Qi2020-jx,Ardagh2019-wb,Shetty2020-sb, berthoumieux2007response, Lemarchand2012-jl, lemarchand2012chemical, berthoumieux2009resonant}. In nonequilibrium and stochastic thermodynamics, similar observations have been made in energy-pumped enzymes, molecular machines, flashing ratchets, and other complicated systems \cite{westerhoff1986enzymes, rozenbaum2004catalytic, qian2005cycle, astumian2001making, yasuda2001resolution,hill1975stochastics, qian1997simple, qian2000simple, astumian2002brownian, reimann2002brownian,astumian1994fluctuation, horowitz2009exact, PhysRevLett.109.203006, astumian2007design, rahav2008directed, sinitsyn2007berry, astumian2007adiabatic, astumian2003adiabatic, astumian2001towards, parrondo1998reversible, astumian2018stochastically, Li1997-xy, reimann1996brownian,PhysRevX.6.021022,liao2021energy}. However, a general theory is still missing to explain the oscillation-pumped phenomenon. So is the universal design principle of catalytic energy landscapes to pump energy from an oscillatory environment to achieve desired performances. 

This letter presents a theoretical framework for oscillation-pumped anomalous catalysis. It explains why, under rapid environmental oscillation, certain catalysts demonstrate anomalous behavior that can never be achieved in stationary environments. The theory geometrically illustrates the underlying mechanisms of a diverse set of anomalous behavior, including inverted (or enhanced) effective thermodynamic driving force, enhanced (or inverted) reaction turnover frequency, and even manipulation of catalytic selectivity. Additionally, the theory can be applied to the rapid oscillation of an arbitrary environmental parameter, including but not limited to temperature, electric field, pressure, light intensity, etc. This work significantly expands the scope of our previous work on temperature-oscillation-induced catalytic inversion \cite{Zhang2022-iq}.

This letter provides a \emph{universal design principle} of oscillation-pumped anomalous performances. Furthermore, this letter introduces a novel control-conjugate landscape description of catalytic reactions, where the reaction rate's dependence on an arbitrary environmental parameter is expressed by exponential forms that resemble the celebrated Arrhenius law. By assuming the existence of a control-conjugate landscape, the design principle can be applied to environmental parameter oscillations of arbitrarily large amplitude. Two types of oscillation-induced performances are demonstrated to exemplify the design principle.

\begin{figure}[h]
    \centering
    \includegraphics{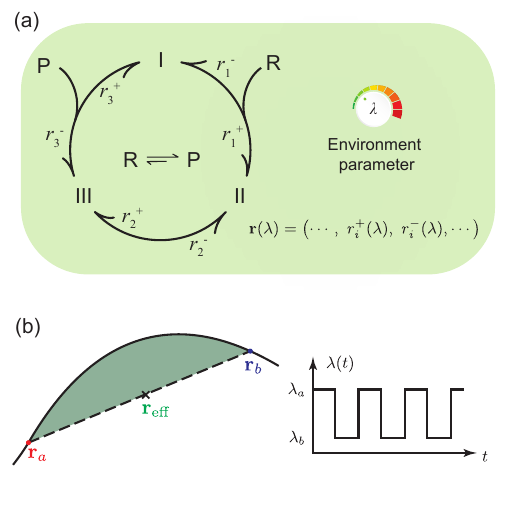}
    \caption{(a) Illustration of a catalyst with $(N=3)$-step catalytic cycle, spontaneously converting reactant R to product P under the controlled environmental condition $\lambda$. At any condition $\lambda$, the catalyst's reaction rates for each step can be sorted into a $2N$-dim vector $\br (\lambda)$. (b) For a continuous range of $\lambda$'s, the vector $\br(\lambda)$ sweep a bend solid black curve in the $2N$-dim space, (i.e., the $\br(\lambda)$-locus). Under rapid environmental oscillations, e.g., the square wave of $\lambda_a$ and $\lambda_b$, the effective rates ${\mathbf r}^\text{eff}$ are achieved at the center point on the straight dash line connecting $\br_a$ and $\br_b$. The green shaded region, i.e., the convex hull of the black curve segment between $\br_a$ and $\br_b$, is the collection of all effective rates that a catalyst can achieve through the arbitrary temporal oscillation within the range $\lambda_a\leq\lambda(t)\leq\lambda_b$. }
    \label{fig:rates}
\end{figure}

\textbf{Theoretical framework for catalysis under rapid oscillation.} 
Consider a general catalytic reaction pathway comprising $N$ intermediate reaction steps with $2N$ reaction rates denoted by $\br=(r^+_1,r^-_{1},r^+_2,r^-_{2},\cdots, r^+_N,r^-_{N})$, where $r^+_i$ and $r^-_i$ denote the forward and backward reaction rates for the $i$-th reaction step. For simplicity,  Fig.~\ref{fig:rates}a illustrates a 3-state catalytic cycle with $N=3$ intermediate steps, converting a reactant R into a product P per clockwise cycle. 
The dynamics of the system can be described by the master equation
\begin{equation}
    \dv{ \vec p }{ t} = \hat R \cdot \vec p 
\end{equation}
where the $\vec p$ is the probability of each intermediate state at a given time and $\hat R$ is a $N\times N$ matrix whose elements are fully determined by the $2N$ reaction rates of the reaction pathway. In general, the catalyst could have multiple cycles and/or involve multiple reactants (and products), and the rates could depend on various environmental parameters. Notice that the reaction rates take into account both the reaction rate constants and the concentrations (activities) of involved chemicals, and the reaction rates are defined per catalyst. 

At a fixed environmental condition, the reaction always reaches a steady state that spontaneously goes in a direction of decreasing free energy ($\Delta G < 0$), which we choose as the forward (clockwise) direction. 
The steady-state probability distribution of the system state $\vec p_{ss}$ is solved by
\begin{equation}
\label{eq:ness}
    \hat R \cdot \vec p_{ss}=0.
\end{equation}
The \emph{ steady-state catalytic performance} of the catalyst can be fully determined by $\vec p_{ss}$ and $\hat R$. Furthermore, due to Eq.~\ref{eq:ness}, the performance can be expressed by a function of the transition rates for all $N$ reactions:
\begin{equation}
\label{eq:fr}
f(\mathbf{r}) = f(r^+_1,r^-_{1},r^+_2,r^-_{2},\cdots, r^+_N,r^-_{N})
\end{equation}

The performance $f(\mathbf{r})$ defined above could take any arbitrary function form depending on the performance of interest. In practice, one may focus on useful performances such as the directional thermodynamic driving force (affinity, see Eq.~\ref{eq:A'}), turnover frequency (net reaction current, see Eq.~\ref{eq:Jr}), and even catalytic selectivity. In stationary conditions, the different designs of catalysts and/or the different choices of stationary environmental conditions are inherently restricted by the nature of the reaction as well as the thermodynamic properties of reactants/products. 
For example, altering the choice of catalysts for a given reaction (fixed reaction free energy) can not change the spontaneous thermodynamic direction reaction. 
Thus, one cannot freely achieve any desired rate vector $\br$ in the $\br$-space merely by redesigning the catalyst. 
As a result, the bounded $\br$-space imposes restraints on the steady-state performance. Below, we discuss how rapid oscillation may break the steady-state performance restraints.

Now let us consider a catalyst driven by a temporally oscillatory environment $\lambda(t)$ with temporal period $\tau$. For finite period $\tau$, the average catalytic performance may depend on the initial state $\vec p(0)$ and the temporal protocol $\lambda(t)$. To circumvent the difficulty and still obtain a universal analytic theory for oscillation-driven catalysis, this work considers a high-frequency limit whose period $\tau$ is much smaller than the fastest transition timescale $\tau \ll \min_{i}(r_i^{-1})$. Yet $\tau$ is still finite, such that one can still consider the rate of an elementary step of the reaction by using the Arrhenius law with a well-defined free energy landscape.

In this rapid oscillation limit, the catalyst reaches a pseudo-stationary state\cite{tagliazucchi2014dissipative,Wang2016-fi,Zhang2022-iq}, which can be solved for as the effective steady state of the effective reaction rates. Here the effective reaction rates are the reaction rates averaged over one oscillation period
\begin{equation}
\label{eq:integ}
    {\mathbf r}^\text{eff} = \frac {1}{\tau} \int_{t_0}^{t_0+\tau}  {\mathbf r}(\lambda(t)) {\rm d} t = \int \rho(\lambda) {\mathbf r}(\lambda) {\rm d} \lambda
\end{equation}
and the time density of $\lambda$ within the period is denoted by $\rho(\lambda)$.
Notice that at the rapid oscillation limit, the choice of time protocol $\lambda(t)$ does not alter the effective rate as long as the time density $\rho(\lambda)$ remains the same.
Here the effective catalytic performance, which is obtained by averaging over a temporal period, can be obtained by plugging in the effective rates ${\mathbf r}^\text{eff}$ into Eq.~\ref{eq:fr} as $f({\mathbf r}^\text{eff})$. In this manuscript, we focus on the design of catalysts rather than optimizing the driving protocol $\lambda(t)$ or effectively $\rho(\lambda)$. In future works, however, given a specific design of a catalyst, one can optimize the performance by looking for the best $\rho(\lambda)$. 

Now we can provide a geometric explanation of why catalysts may show novel performances under rapid environmental oscillation. Depending on the specific system and the performance of interest, whose steady-state performances may be restricted by $f\leq c$ (or $f\geq c$), oscillation-induced effective rate ${\mathbf r}^\text{eff}$ may break the steady-state restraints by enhancement (or suppression) of the performance and achieve $f>c$ (or $f<c$).
In Fig.~\ref{fig:rates}b, we define the $2N$-dim $\br$-space comprising of the $2N$ reaction rates. We sketch a black curve to represent the family of reaction rates $\br(\lambda)$'s that a catalyst can achieve at any stationary environmental condition $\lambda$. We will denote the black curve as the \emph {stationary-rate locus} or $\br(\lambda)$-locus. 
When environmental condition rapidly oscillates according to an arbitrary periodic protocol, as argued in Eq.~\ref{eq:integ}, the catalyst can achieve a new effective rate matrix, i.e., any point ${\mathbf r}^\text{eff}$ within the convex hull of the corresponding segment of the $\br(\lambda)$-locus bounded by the range of $\lambda(t)$. The convex hull is illustrated as a green region in Fig.~\ref{fig:rates}b. 

\begin{figure*}[ht]
    \centering
    \includegraphics{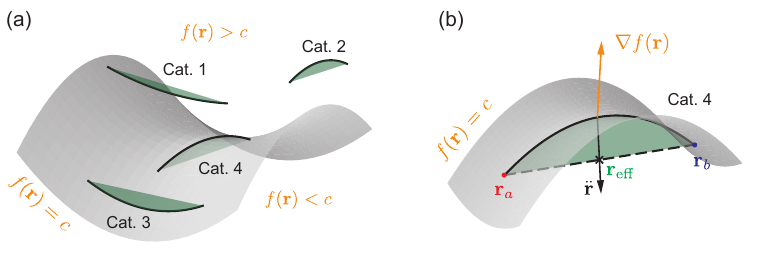}
    \caption{The geometric illustration of designing oscillation-induced catalysis in the $2N$-dim reaction rates $\br$ space. (a) 
    The gray surface represents the iso-performance hyper-surface $f(\br)=c$. The four black curves each represent a candidate catalyst, and they can only achieve steady-state performance restricted by $f(\br)\geq 0$. The green-shaded convex hull of each black curve indicates the set of effective rates that each catalyst can achieve under environmental oscillation. Notice that the convex hull of Cat.~4 breaks the restriction and can achieve effective performance $f(\br)<c$. (b) Geometric analysis of Cat.~4 by illustrating its acceleration vector, $\ddot \br$ (bending direction), and the gradient of performance function, $\nabla f(\br)$. When the two vectors point in opposite directions, the effective rates achieved by oscillation could exhibit performance suppression $f<c$.
    }

    \label{fig:geo}
\end{figure*}

If the catalyst's $\br(\lambda)$-locus is bent in the $\br$-space, its convex hull introduces new points beyond the $\br(\lambda)$-locus. As a result, the catalyst may achieve effective rates ${\mathbf r}^\text{eff}$ and performances that are never accessible in any stationary environment. 
This can be seen by first recognizing that the steady-state performance, regardless of the choices or designs of catalysts, may be restricted to a range (e.g., $f(\mathbf r)\geq c$ or $f(\mathbf r)\leq c$). By reaching effective rates from the convex hull, the effective performance of a catalyst, $f(\mathbf r^\text{eff})$, may exceed the steady-state restraints if $f(\mathbf r^\text{eff})<c$, or $f(\mathbf r^\text{eff})>c$. Notice that the steady-state performance restraints may either be an upper bound or a lower bound, depending on the specifics of the systems and the performance of interest. In Fig.~\ref{fig:geo}, we illustrate the argument by choosing a convention where the steady-state performance is restricted by $f(\mathbf r)\geq c$, and oscillation-pumped catalyst achieves an effective performance beyond the restriction $f(\mathbf r^\text{eff})<c$.

For illustrative purposes, Fig.~\ref{fig:rates}b demonstrates a simple binary oscillatory environment with half a period in $\lambda_\text{a}=\lambda_0-\Delta \lambda/2$ and another half in $\lambda_\text{b}=\lambda_0+\Delta \lambda/2$. In the rapid oscillation limit, each reaction step has $1/2$ chance to occur with one rate $r_{i,\text{a}}$, and $1/2$ chance to occur with another rate $r_{i,\text{b}}$. Thus, the effective rates for all reactions are the simple arithmetic mean of the rates of the two environmental conditions:
\begin{equation}
\label{eq:eff-rate}
    {\mathbf r}^\text{eff} = \frac{{\mathbf r}(\lambda_{\text{a}}) + {\mathbf r}(\lambda_{\text{b}} )}{2}
\end{equation}
which typically differs from ${\mathbf r}(\lambda_0)$ or any ${\mathbf r}(\lambda)$. One could alter the effective rate by changing the fraction of time spent in $\lambda_{\text{a}}$ and $\lambda_{\text{b}}$.

\textbf{Universal Design Principle}
In a stationary environment, the performance constraints may take the form of $f(\br)\geq c$ or $f(\br)\leq c$ for all $\br(\lambda)$, and here we focus on a geometric design principle of oscillation-driven catalysts to achieve effective rates ${\mathbf r}^\text{eff}$ and effective performances beyond the stationary-environment restraints, i.e., $f({\mathbf r}^\text{eff}) <c$ or $f({\mathbf r}^\text{eff}) >c$.

Recall, in Fig.~\ref{fig:rates}b, that the rates of a catalyst within a stationary environmental condition $\lambda$ are collectively expressed by a vector $\br(\lambda)$. For a range of stationary environmental conditions, $\lambda$'s, the rate vector sweeps through a curve, i.e., the $\br(\lambda)$-locus. For the same range of $\lambda$'s, each design of a catalyst can be represented by a curve, and the endeavor to design an optimal catalyst with desired oscillation-driven performance is equivalent to looking for an optimal curve $\br(\lambda)$-locus. In Fig.~\ref{fig:geo}a, four candidate catalysts are illustrated by four $\br(\lambda)$-locus in the $\br$-space. Notice that under any stationary condition, the performances of all candidate catalysts may be restricted by the thermodynamic and kinetic rules. In the illustration, this restriction, illustrated as $f(\br)>c$, can be seen as all $\br(\lambda)$-locus of all 4 catalysts are restricted an iso-performance manifold (i.e., $f(\br)>c$ for all $\br(\lambda)$'s. The ability for a catalyst to break the stationary-environment restraint lies in the location and the bending direction of the $\br(\lambda)$-locus of each catalyst. In the illustrated example, only the ``Cat.~4'' can achieve $f({\mathbf r}^\text{eff})<c$, as its locus sits on/above the iso-performance surface ($f(\br)= c$) and the locus bends toward the lower side $f(\br)<c$.

Given the above geometric analysis, the universal design principle of a catalyst for achieving higher (or lower) effective performance value $f({\mathbf r}^\text{eff})$, can be expressed by an objective function $\mathcal{C}^{[f]}$: 
\begin{equation}
\label{eq:CCC}
\begin{split}
    \mathcal{C}^{[f]} & \equiv  \ddot{\mathbf{r}}  \cdot  \nabla_{\mathbf{r}} f \\
    &= \sum_{i=1}^{N} \left ( \pdv{f}{r^-_{i}} \cdot \dv[2]{r^-_{i}}{\lambda}+\pdv{f}{r^+_{i}} \cdot \dv[2]{r^+_{i}}{\lambda}\right )
\end{split}
\end{equation}
whose maximization (or minimization) serves as the universal \emph{design criterion} for optimal catalysts. This objective function $\mathcal{C}^{[f]}$ takes into consideration both the geometric bending of the $\br(\lambda)$-locus and the gradient vector of the desired performance $f({\mathbf r})$. The bending direction and amplitude of $\br(\lambda)$-locus can be locally described by the acceleration vector, 
\begin{equation}
    \ddot{\br}\equiv \dd^2{\mathbf{r}}/\dd{\lambda}^2 = \left(\cdots,~ \frac{{\rm d}^2r^+_i}{{\rm d}\lambda^2},~\frac{{\rm d}^2r^-_i}{{\rm d}\lambda^2 },~\cdots \right)
\end{equation}
The gradient vector of the performance is
\begin{equation}
    \nabla _{\mathbf{r}} f \equiv  \left(\cdots, ~\frac{\partial f}{\partial r^+_i},~\frac{\partial f}{\partial r^-_i},~\cdots\right)
\end{equation}
which captures the direction where the performance increases most rapidly.

The geometric argument behind the objective function (Eq.~\ref{eq:CCC}) is that one can predict the oscillation-induced performance change (i.e., higher or lower values of $f$) by the geometric alignment between the acceleration vector $\ddot{\br}$ and the (negative) gradient vector $\nabla _{\mathbf{r}} f$. Specifically, when the two vectors ($\ddot{\br}$ and $\nabla _{\mathbf{r}} f$) are aligned, the oscillation-driven catalyst achieves a higher value of performance $f$; when the vectors are anti-aligned, the oscillation-driven catalyst achieves a lower value of performance $f$; and when the vectors are perpendicular to each other, the catalyst's performance is insensitive to the environmental oscillation. (See Fig.~\ref{fig:geo}b for an example of anti-aligned vectors $\ddot{\br}$ and $\nabla _{\mathbf{r}} f$.)

Notice that at small amplitude oscillation of $\lambda$, this geometry-inspired criterion $\mathcal{C}^{[f]}$ rigorously quantifies the oscillation-induced performance deviation. Moreover, as shown by the following geometrical argument, the criterion still applies well even for large-amplitude $\lambda$ oscillations.

\begin{figure}
    \centering
    \includegraphics{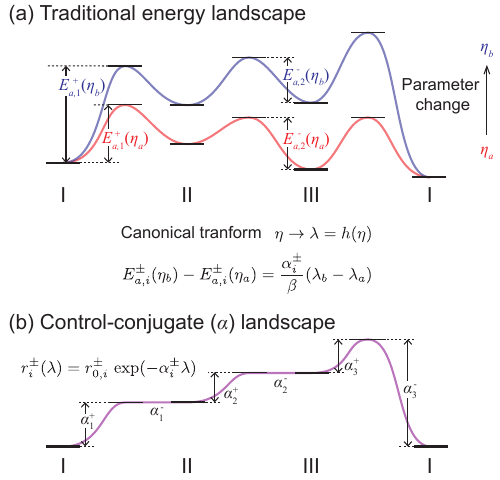}
    \caption{Proposed control-conjugate landscape as a concise representation of a family of energy landscapes tuned by environmental condition $\eta$. (a) Two energy landscapes of the reaction coordinate of a $3$-state catalytic cycle are illustrated for two different environmental conditions $\eta_a$ and $\eta_b$. The canonical environmental condition $\lambda = h(\eta)$ is defined so that the activation energies linearly depend on $\lambda$. (b) The reaction rates' dependence on the environmental control $\lambda$ is collectively represented by the control conjugate landscape. A $\alpha^{\pm}_i >0$ (or $\alpha^{\pm}_i <0$) indicates that the corresponding reaction rate exponentially decreases (or increases) with $\lambda $. A flat ``barrier'', $\alpha^{\pm}_i=0$, indicates that the rate of the corresponding reaction step is independent of the environmental parameter.}
    \label{fig:landscape}
\end{figure}

\textbf{Design Principle of Control Conjugate Landscape}
In general, the applicability of the universal design principle is limited by two complications. Firstly, the theory does not assume a specific representation of the physical quantity to capture the environmental condition. Secondly, encoding the catalyst's response to environmental conditions in a single energy landscape is impractical. Instead, the catalyst's reaction kinetics under a spectrum of control parameters should be represented by ``energy seascape'', i.e., a family of energy landscapes parameterized by the environmental condition.\cite{dill2012protein,chen2000rna} To resolve these difficulties, below we define a canonical control variable $\lambda$, and introduce the control-conjugate landscape to capture the whole energy seascape or the family of reaction rates $\textbf{r}(\lambda)$ parameterized by $\lambda$. The control-conjugate landscape is inspired by Arrhenius's law, where the control conjugate landscape is analogous to the energy landscape, and the canonical control parameter $\lambda$ is analogous to the inverse temperature $\beta$. The task of designing a novel catalyst becomes designing the control-conjugate landscape.

Consider that the environmental system is directly controlled by physical quantity $\eta$, which is in a practical representation. Here the non-canonical $\eta$ could represent temperature, system volume, light intensity, etc., with arbitrarily chosen unit systems. Each reaction step's rate can be expressed by a non-canonical function of $\eta$, and may assume complicated forms. For instance, if $\eta\equiv T$) is chosen as the temperature in the Kelvin representation, then according to Arrhenius law, reaction rates can be written as
\begin{equation}
\label{eq:ah}
r^\pm_{i} = A_i \exp\left (\frac{-E^{\pm}_{a,i}}{R T}\right ) =   A_i \exp\left ( - \beta E^{\pm}_{a,i} \right)
\end{equation}
where $A_i$ is a constant independent of temperature, and $E^{\pm}_{a,i}$ is the activation energy that can be obtained from the energy landscape (Fig.~\ref{fig:rates}b). 
Notice that the Arrhenius rate may assume more complicated forms if the temperature is denoted by the Celsius or Fahrenheit representation. However, the rate is represented in a canonical form at the second equality where we have adopted the canonical variable (``inverse temperature''), $\beta = 1/(R T)$; the reaction rates assume an exponential dependence on the canonical control parameter.

In more general cases, i.e., where the changing environmental condition could be system volume, light intensity, etc., we define the ``canonical control parameter'' $\lambda$ and the corresponding control-conjugate landscape such that reaction rates follow a canonical form, i.e., exponentially depend on $\lambda$. 
Consider the activation energies $ E^{\pm}_{a,i}(\eta)$ as arbitrary functions of the non-canonical environmental parameter $\eta$, then the reaction rates can be expressed as functions of $\eta$
\begin{equation}
    r^\pm _{i} =  A_i \exp \left (-\beta E^{\pm}_{a,i}(\eta)\right )  
\end{equation}
We can define a canonical transformation from $\eta$ to $\lambda=h(\eta)$ if, for any reaction step and any choices of $\eta_a$ and $\eta_b$ within the control range, the dimensionless activation energy linearly depend on $\lambda$:
\begin{equation}
    \label{eq:linear}
    \beta E^{\pm}_{a,i}(\eta)=\alpha^{\pm}_i \lambda + {\rm{const.}}
\end{equation}
In this case, the rates' dependence on the canonical control parameter $\lambda$ follows the canonical exponential form:
\begin{equation}
\label{eq:rate}
    r^{\pm}_{i} = r^{\pm}_{0,i}\exp(-\alpha^{\pm}_{i} \lambda)
\end{equation}
where $\lambda$ is analogous to the inverse temperature $\beta$, and $\alpha^{\pm}_{i}$ is analogous to the activation energy in the Arrhenius law. Here we shall refer to $\alpha^{\pm}_{i}$ as the \emph{control-conjugate activation}, and thus we can define a control-conjugate landscape (see Fig.~\ref{fig:landscape}b).
In Fig.~\ref{fig:landscape}, we illustrate the linear dependence in Eq.~\ref{eq:linear} by choosing two environmental conditions, and we determine the control conjugate activation $\alpha^{\pm}_i$ according to
\begin{equation}
          \beta E^{\pm}_{a,i}(\eta_b)- \beta E^{\pm}_{a,i}(\eta_a)=\alpha^{\pm}_i(\lambda_b-\lambda_a).
 \end{equation}

Notice that the conjugate energy landscape framework is general enough to describe catalysts with selective environmental responses (i.e., when the environmental control only impacts the rate of a few reaction steps). In other words, if the transition rate $r_j$ is independent of the external control $\lambda$, then $\alpha_{j}= 0$. 

The general design principle (Eq.~\ref{eq:CCC}) can be significantly simplified and generalized if one can obtain the canonical representation $\lambda$ and the control conjugate landscape. I.e., designing a catalyst with enhanced performance $f$ becomes designing the control conjugate activation such the following objective function is optimized:
\begin{equation}
\label{eq:CCCsimp}
    \mathcal{C}^{[f]} = \sum_{i=1}^{N} \left (r^-_{i}(\alpha_i^-)^2  \pdv{f}{r^-_{i}}  +  r^+_{i}(\alpha_i^+)^2 \pdv{f}{r^+_{i}} \right )
\end{equation}
Moreover, the canonical form of rates (Eq.~\ref{eq:rate}) allows one to apply the design principle to large amplitude oscillations. 
This is geometrically justified because the $\br(\lambda)$ locus bears a well-behaving non-exotic shape where the local bending direction is representative of its global bending direction -- at the logarithm-scale representation of the $\br$-space (see Fig.~\ref{fig:geo}b), the locus reduces to a straight line, with the angle the line defined by the control-conjugate activations (see Eq.~\ref{eq:rate}). One can verify the result by applying it to design the catalytic inversion of spontaneous reaction under rapid temperature oscillation, where the general principle reduces to the result obtained in our previous work \cite{Zhang2022-iq}.

\textbf{Examples. }
The universal design criterion $\mathcal{C}^{[f]}$ can guide the design of oscillation-driven catalysts with desired performance $f$. 
For illustrative purposes, this paper exemplifies the design of two interesting anomalous behavior: (1) oscillation-induced inversion of spontaneous reaction direction, and (2) oscillation-enhanced high turnover frequency without utilizing low activation barriers as required by the traditional theory. Notice that the design principle applies to any desired interest performance $f$, not limited to these two examples. In all cases, the anomalous behavior is sustained by the energy harnessed from the oscillatory environment.

\begin{figure*}[ht]
    \centering
    \includegraphics{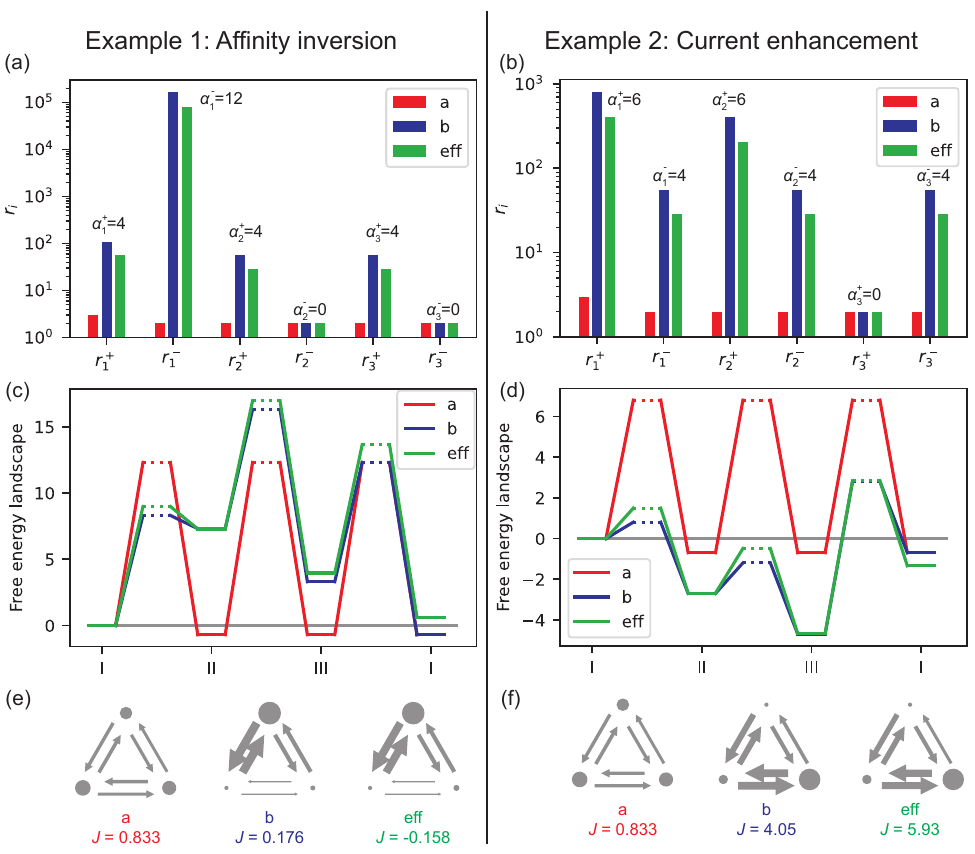}
    \caption{Demonstrations of example-1 (a,c,e) and example-2 (b,d,f). Dynamics at two stationary environments ($\lambda_a$ and $\lambda_b$) are shown in red and blue; the oscillation-induced effective dynamics are in green.
    (a,b) illustrate stationary reaction rates and effective rates (Eq.~\ref{eq:eff-rate}). (c,d) illustrate two dimensionless stationery free-energy landscapes $G/RT$ and the effective landscape.
    The tilt of the (effective) landscape, i.e., the height difference between two ends, which equals $-\tilde{\mathcal{A}}$, indicates the (effective) spontaneous direction. 
    (e,f) illustrates the detailed steady-state currents of each transition (thickness of arrows) and the steady-state probability of intermediate states (size of disks). $J$ represents the net reaction currents.
    For both applications, we set $r^+_{1}(\lambda_a)=2$, and $r^-_{1}(\lambda_a)=r^+_{2}(\lambda_a)=r^-_{2}(\lambda_a)=r^+_{3}(\lambda_a)=r^-_{3}(\lambda_a)=1$ for $\lambda_a=0$. The rates for $\lambda_b=-1$ can be calculated from $r^{\pm}_{i}(\lambda_a)$'s and $\alpha^{\pm}_i$'s. For both $\lambda_a$ and $\lambda_b$, $\Delta G_\text{rxn}/RT = -\tilde{\mathcal{A}} =-\ln 2$. }
    \label{Fig:app}
\end{figure*}

{\bf Example 1 -- Inverting Thermodynamic Affinity}. Here demonstrated is the optimal catalytic energy landscape for inverting the spontaneous direction of the reaction ($\tilde{\mathcal{A}}$). Such inversion is particularly interesting as the catalyst harnesses energy from the external environment and stores the energy by converting low-free energy products into high-free-energy reactants. We choose the performance $f({\mathbf r})$ to represent the thermodynamic spontaneity (i.e., dimensionless thermodynamic affinity)
\begin{equation}
\label{eq:A'}
    \tilde{\mathcal{A}}({\mathbf r}) = \sum_{i=1}^N \log \frac{r^+_i}{r^-_{i}}
\end{equation}
which is simply related to the reaction Gibbs free energy change via $\tilde{\mathcal{A}}=-\beta \Delta G$.
To maximally invert the spontaneous reaction, one needs to find a catalyst with a very negative $\mathcal{C}^{[\tilde{\mathcal{A}}]}$. In other words, the universal principle Eq.~\ref{eq:CCC} for designing strong catalytic inversion of a spontaneous reaction becomes
\begin{equation}
\label{eq:principle1}
    \mathcal{C}^{[\tilde{\mathcal{A}}]} = \sum_{i=1}^N (\alpha^+_i)^2- \sum_{i=1}^N  (\alpha^-_{i})^2
\end{equation}
which is a simple quadratic function that depends only on $\alpha_i$'s, each step's activation level conjugates to the control $\lambda$. 
In conclusion, a designed catalyst with very negative $\mathcal{C}^{[\tilde{\mathcal{A}}]}$ tends to strongly invert the spontaneous reaction direction under rapid oscillatory control. 

The design principle Eq.~\ref{eq:principle1} can be experimentally verified. Here we illustrate an optimal catalyst obtained by the principle under a few constraints: To prevent the control parameter $\lambda$ from directly impacting the spontaneity $\tilde{\mathcal{A}}$, we require that 
\begin{equation}
\label{eq:const}
    \dv{\tilde{\mathcal{A}}}{\lambda} = \sum_{i=1}^N (\alpha^+_i-\alpha^-_{i}) = 0
\end{equation}
and thus, any observed inversion of the reaction's effective direction is purely a result of the catalytic response to rapid environment oscillation. 
By suppressing $\sum_{i=1}^N (\alpha^+_i)^2$ and increasing $\sum_{i=1}^N (\alpha^-_i)^2$ under the constraint Eq.~\ref{eq:const}, the optimal reaction-inverting catalyst for $N=3$ can be designed as
\begin{align}
\label{eq:+inv}
\begin{split}
    (\alpha^+_{1}, \alpha^+_{2}, \alpha^+_{3}) &= (\frac{1}{3}\alpha, \frac{1}{3}\alpha, \frac{1}{3}\alpha) \\
    (\alpha^-_{1}, \alpha^-_{2}, \alpha^-_{3}) &= (\alpha, 0, 0)
\end{split}
\end{align}
This design is illustrated in Fig.~\ref{Fig:app}a, where the two sets of reaction rates respectively correspond to environmental conditions $\lambda_a=0$ (red) and $\lambda_b=-1$ (blue), and the effective rates due to rapid oscillation are shown in green colored bars.

Shown in Fig.~\ref{Fig:app}c, the effective reaction energy landscape for this catalyst has an inverted tilted direction compared to the stationary-condition landscapes. In other words, the effective affinity has an opposite sign of the affinity of the stationary environmental condition, leading to the reaction direction inversion. As illustrated in Fig.~\ref{Fig:app}e ($\alpha=12$), under constant environment $\lambda_a$ (or $\lambda_b$), the steady-state current is $J_a=0.833$ (or $J_b=0.176$), whereas a rapid oscillation between these two conditions leads to an average current $J_\text{eff}=-0.158$. Thus the catalyst actively harnesses energy from the environment to drive the reaction against its spontaneous direction. A special case of catalytic thermal engines under temperature oscillation is demonstrated in reference \cite{Zhang2022-iq}.

{\bf Example 2 -- Enhancing reaction TOF (current)}. Here we illustrate the design principle for optimal enhancement of reaction current $J$, which is equivalent to turnover frequency (TOF). The TOF is actively enhanced by energy harnessed from the environment, and cannot be explained by traditional catalysis theory, which focuses on lowering activation barriers. 
In this case, we choose performance $f({\mathbf r})$ as the (pseudo-)stationary current $J({\mathbf r})$, which can be analytically solved for arbitrary catalytic cycles (see reference \cite{Hill1988-tp}). For a $N=3$-step catalytic cycle, the current is
\begin{align}
    J({\mathbf r}) &= \frac{u({\mathbf r})}{v({\mathbf r})}\label{eq:Jr}\\
    u({\mathbf r}) &= r^+_1 r^+_2 r^+_3 - r^-_{1} r^-_{2} r^-_{3} \label{eq:ur}\\
    \begin{split}
    v({\mathbf r}) &= r^+_1 r^+_2 + r^+_2 r^+_3 + r^+_3 r^+_1\\
    &\ + r^-_{1} r^-_{2} + r^-_{2} r^-_{3} + r^-_{3} r^-_{1} \\
    &\ + r^+_{1} r^-_{2} + r^+_{2} r^-_{3} + r^+_{3} r^-_{1}
    \end{split}
\end{align}
Plugging into the universal principle, Eq.~\ref{eq:CCC}, we find the design criteria for enhancing TOF as to increase the value of
\begin{equation}
\label{eq:enhanceJ}
    \mathcal{C}^{[J]} = J (\frac{u_2}{u} - \frac{v_2}{v})
\end{equation}
where
\begin{align}
\begin{split}
    u_2 &= \left((\alpha^+_1)^2+(\alpha^+_2)^2+(\alpha^+_3)^2\right ) r^+_1 r^+_2 r^+_3 \\
    &\ - \left((\alpha^-_{1})^2+(\alpha^-_{2})^2+(\alpha^-_{3})^2\right) r^-_{1} r^-_{2} r^-_{3}  \\
\end{split} \\
\begin{split}
    v_2 &=\left( (\alpha^+_1)^2+(\alpha^+_2)^2\right) r^+_1 r^+_2 +\left( (\alpha^+_2)^2+(\alpha^+_3)^2\right) r^+_2 r^+_3 \\
    &\ +\left( (\alpha^+_3)^2+(\alpha^+_1)^2\right) r^+_3 r^+_1 +\left( (\alpha^-_{1})^2+(\alpha^-_{2})^2\right) r^-_{1} r^-_{2} \\
    &\ +\left( (\alpha^-_{2})^2+(\alpha^-_{3})^2\right) r^-_{2} r^-_{3} +\left( (\alpha^-_{3})^2+(\alpha^-_{1})^2\right) r^-_{3} r^-_{1} \\
    &\ +\left( (\alpha^+_{1})^2+(\alpha^-_{2})^2\right) r^+_{1} r^-_{2} +\left( (\alpha^+_{2})^2+(\alpha^-_{3})^2\right) r^+_{2} r^-_{3} \\
    &\ +\left( (\alpha^+_{3})^2+(\alpha^-_{1})^2\right) r^+_{3} r^-_{1}
\end{split} 
\end{align}
Here Eq.~\ref{eq:enhanceJ}, as a function of $r_i$'s as well as $\alpha_i$'s can be directly used as a design principle for the numerical search of the optimal design of oscillation-enhanced catalyst. Moreover, by observing this equation, we can further obtain approximated design rules leading to a simple working example of a prominent current-enhancing catalyst under environment oscillation.
Notice that ${v_2}/{v}$ in Eq.~\ref{eq:enhanceJ} is always positive. Thus to enhance the performance, or achieve a large positive $\mathcal{C}^{[J]}$, we need to find a catalyst of both large ${u_2}/{u}$ and large $J$.
To increase ${u_2}/{u}$, one can enhance $\sum_{i=1}^N (\alpha^+_i)^2$ and suppress $\sum_{i=1}^N (\alpha^-_i)^2$.
Combining these considerations, we obtain an intuitive catalytic landscape characterized by 
\begin{align}
\label{eq:enh}
\begin{split}
    (\alpha^+_{1}, \alpha^+_{2}, \alpha^+_{3}) &= (\alpha, \alpha, 0) \\
    (\alpha^-_{1}, \alpha^-_{2}, \alpha^-_{3}) &= (\frac{2}{3}\alpha, \frac{2}{3}\alpha, \frac{2}{3}\alpha)
\end{split}
\end{align}
An example of choosing $\alpha=6$ is summarized in Fig.~\ref{Fig:app}b,d,f.
The reaction rates for $\lambda_a$, $\lambda_b$, and the effective rates are shown in Fig.~\ref{Fig:app}b. The corresponding dimensionless free energy landscapes are shown in Fig.~\ref{Fig:app}d. In Fig.~\ref{Fig:app}f, we verify that the effective reaction TOF (current) reaches $J_{\text{eff}}=5.93$ under rapid oscillation, which is significantly greater than the stationary current of both conditions ($J_a=0.833$, and $J_b=4.05$).

This paper describes a general non-equilibrium theory framework predicting what catalysts could demonstrate anomalous performance under rapid environment oscillation and why. The theory generally applies to catalysts with an arbitrary number of intermediate steps $N$, and proposes a convenient control-conjugate landscape to describe the catalyst's response to different environmental conditions $\lambda$'s. Furthermore, for any oscillatory control parameter, the theory provides a geometric design principle for engineering optimal catalysts with a wide range of oscillation-enhanced performances $f$, such as inverting spontaneous reaction direction or enhancing reaction rate, both empowered by the dissipation of energy. Immediate applications following the two examples listed above can be found in oscillation-enhanced redox reactions and electrolysis, where one can empower a reaction by low-magnitude AC voltages rather than high-magnitude DC voltages. For instance, traditional DC electrolysis is subject to the thermodynamic limit (theoretical cell voltage) plus an extra kinetic limit (overpotential). With novel catalysts, the reaction could occur under oscillatory voltage, even if it is below the theoretical cell voltage. In practice, the weak voltage can avoid undesired side reactions such as electrolysis of the solvent. 

Beyond these two examples, by defining other types of performances $f$, one can achieve more complex and useful performances such as selectivity manipulation. For instance, consider a complex catalytic network with multiple loops, where one loop produces desired products while other loops produce by-products. We can design oscillation-responsiveness to each loop and define the performance $f$ as selectivity, which is the ratio between the net current of the desired-product loop and the total net current among all loops. This allows us to manipulate the catalytic selectivity by adjusting the external oscillatory stimulation.

The fundamental geometric intuition in the design space to explain oscillation-induced anomalous performances is general and not restricted by the simplicity of control parameter $\lambda$ and the control-conjugate landscape. As represented in the $\br$-space, a catalyst under any stationary environment is described by a point $\br(\lambda)$. When the environment rapidly oscillates, the catalyst's kinetic rates are effectively represented by an average point $\br_{\text{eff}}$ in the same $\br$-space, allowing for a new performance that can not be achieved at any stationary condition. Even if the environmental control is complicated (e.g., when reaction rates cannot be represented by a control-conjugate landscape or when the environmental control involves multi-dimensional $\vec \lambda$), one can still utilize the geometric intuition to obtain design principles for desired oscillation-induced performances.

In the future, this work opens up the opportunity to design catalysts beyond the stationary theory, where catalysts are not simply designed by lowering activation barriers as required by the traditional theory. Beyond what is discussed in this paper, one can also find design principles in terms of the optimal control protocols of $\lambda(t)$. This can potentially allow catalysts to function as environmental sensors, producing signaling molecules only when the environment oscillates according to specific temporal patterns.

\begin{acknowledgement}
We acknowledge the fund from the National Science Foundation Grant DMR-2145256. 
We appreciate helpful discussions and suggestions for the manuscript from Hong Qian, Zhixin Lu, Yosuke Kanai, Jahan Dawlaty, Ruicheng Bao, Shiling Liang, and Jim Cahoon.
\end{acknowledgement}

\bibliography{ref}

\end{document}